\documentclass{mn2e}
\input{epsf}

\title[Andromeda II as a merger remnant]
{Andromeda II as a merger remnant}
\author[E. L. {\L}okas et al.]
    {Ewa L. {\L}okas$^{1}$, Ivana Ebrov\'{a}$^{2,3}$,  Andr\'{e}s del Pino$^{4}$ and Marcin Semczuk$^{5}$
    \\
    $^1$Nicolaus Copernicus Astronomical Center, Bartycka 18, PL-00-716 Warsaw, Poland\\
    $^2$Astronomical Institute, Academy of Sciences of the Czech Republic, Bo\v{c}n\'{i} II 1401/1a, CZ-141 00 Prague,
	Czech Republic\\
    $^3$Institute of Physics, Academy of Sciences of the Czech Republic, Na Slovance 1999/2 CZ-182 21 Prague,
	Czech Republic\\
    $^4$Instituto de Astrof\'{i}sica de Canarias, Calle V\'{i}a L\'{a}ctea s/n, E-38200 La Laguna, Tenerife,
    Canary Islands, Spain\\
    $^5$Astronomical Observatory of the Jagiellonian University, Orla 171, PL-30-244 Cracow, Poland}

\begin{document}

\maketitle

\begin{abstract}
Using $N$-body simulations we study the origin of prolate rotation
observed in the kinematic data for Andromeda II,
a dSph satellite of M31. We propose an evolutionary model for the origin of And II involving a merger between two
disky dwarf galaxies with different disk scale lengths. The dwarfs are placed on a radial orbit towards each
other with their angular momenta inclined by 90 deg. The merger remnant forms a stable triaxial galaxy with rotation
only around the longest axis whose origin is naturally explained as due to the symmetry of the initial configuration
and the conservation of angular momentum components along the direction of the merger. The stars originating from the
two dwarfs show significantly different surface density profiles while having very similar kinematics as required by
the data. We also study an alternative scenario for the formation of And II, via tidal stirring of a disky dwarf.
While intrinsic rotation occurs naturally in this model as a remnant of the initial rotation of the disk, it is
mostly around the shortest axis of the stellar component. We conclude that the velocity distribution in And II is
much more naturally explained by a scenario involving a past merger.
\end{abstract}


\vspace{-1cm}

\section{Introduction}

Dwarf spheroidal (dSph) galaxies of the Local Group provide unique laboratory for testing scenarios of structure
formation in the Universe. Due to their proximity, observations allow us to resolve single stars in these systems
and measure their detailed kinematics. Such measurements revealed that random motions of
stars dominate in dSphs and they are therefore considered to be pressure-supported systems. While some evidence
for streaming motions in these systems exists (see e.g. a summary in {\L}okas et al. 2011) it is not very convincing
except for a recent study of Andromeda II (And II), a dSph satellite of M31, by Ho et al. (2012). They have demonstrated
that the galaxy displays a strong rotation signal, at its maximum comparable to the central velocity dispersion.
What is even more surprising, the rotation appears to be around the major axis of the stellar component.
The reality of this streaming motion was recently confirmed by Amorisco, Evans \& van de Ven (2014) by reanalysis
of the same data that also suggested the presence of substructure.

One of the commonly accepted scenarios for the formation of dwarf spheroidal galaxies in the Local Group
proposes that they were formed from disky dwarf progenitors accreted by a bigger galaxy such as
the Milky Way or M31. While orbiting their host, the dwarfs are affected by tidal forces from the big
galaxy and in addition to mass loss their stellar components gradually transform from disks into spheroids
and the initial rotation is replaced by random motions of the stars. The ordered motion (rotation) disappears
completely only for dwarfs evolving on very tight orbits, the scenario thus offers a natural explanation for
the presence of rotation in dSphs as the remnant of their past (for details see Mayer et al. 2001; Klimentowski
et al. 2009; Kazantzidis et al. 2011a; {\L}okas et al. 2011, 2012).

Another possible way to form a dSph galaxy is by a major merger of two late-type dwarfs. While such events are believed
to be rare in the environments such as the Local Group, they do occur as was demonstrated by Klimentowski et al.
(2010), although rather early on and at the outskirts of the Group. Selected merger events identified in this
study were resimulated with a higher resolution and including disky stellar components by
Kazantzidis et al. (2011b) and it was shown that in majority of cases they lead to the formation of galaxies with
properties very similar to dSphs of the Local Group, again with some remnant rotation.

Inspired by these studies, we propose an evolutionary model for And II in which it originates as a result of a merger
between two similar late-type dwarf galaxies. We demonstrate that this scenario explains in a natural way the amount and
type of rotation in And II. In particular, the prolate rotation (around the major axis of the remnant)
occurs as a result of the conservation of angular
momentum along the direction of the merger. We also show that similar effect is difficult to obtain in the tidal
stirring scenario where the remnant rotation is mostly around the minor axis of the stellar component.

\vspace{-0.5cm}

\section{Simulation of the merger}

\begin{figure}
\begin{center}
    \leavevmode
    \epsfxsize=8cm
    \epsfbox[0 10 435 275]{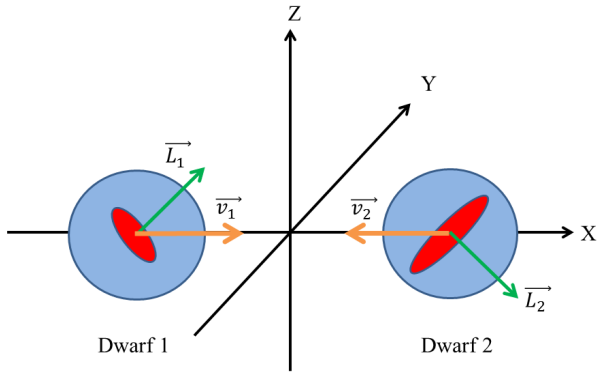}
\end{center}
\caption{The initial simulation setup.}
\label{merger}
\end{figure}

The initial simulation setup is presented in Figure~\ref{merger}. Two similar dwarf galaxies were placed at the
distance of 25 kpc from the centre of the coordinate system on the X axis of the simulation box. They were assigned
equal velocities of 8 km s$^{-1}$ toward the centre, i.e. again along the X axis. The disks of the two dwarfs were
oriented so that their angular momentum vectors were confined to the XZ plane and inclined by $\pm 45$ deg to the
XY plane.

While such a configuration may seem rather special, we note that And II is the only dSph in the Local
Group in which the unusual property of prolate rotation was detected so the required initial conditions are probably
not very common. However, Kazantzidis et al. (2011b) were able to identify (among a few hundred of subhaloes
studied in the Local Group simulation)
at least eight merger events with progenitors of comparable mass, a variety of
angular momentum orientations and on approximately radial orbits. Most of these mergers led to the formation of
objects with properties akin in terms of shape and the amount of rotation to those of Local Group dSph galaxies.

Each of the two dwarf galaxies was initially composed of an exponential disk embedded in an NFW dark matter halo.
The first dwarf (Dwarf 1) had a dark halo of mass $M_{\rm h} = 10^9$ M$_{\odot}$ and concentration $c=20$. Its disk had a
mass $M_{\rm d} = 2 \times 10^7$ M$_{\odot}$, exponential scale-length $R_{\rm d} = 0.41$ kpc and thickness
$z_{\rm d}/R_{\rm d} = 0.2$. Thus the dwarf galaxy model used was the same as the one recently studied in the
simulation of tidal stirring in {\L}okas et al. (2014).
The second dwarf (Dwarf 2) differed from the first only by the disk scale-length which
was assumed to be $R_{\rm d} = 0.67$ kpc and thickness (since we kept $z_{\rm d}/R_{\rm d} = 0.2$).
This larger scale-length is well within the range expected for haloes of this mass (see Kazantzidis
et al. 2011a).

The $N$-body realizations of the dwarfs were
generated via procedures described in Widrow \& Dubinski (2005) and Widrow, Pym \& Dubinski (2008).
We used $2 \times 10^5$ particles in each component of each dwarf (i.e. $8 \times 10^5$ particles total).
The adopted softening scales were $\epsilon_{\rm d} = 0.02$ kpc and $\epsilon_{\rm h} = 0.06$ kpc for the
disk and halo respectively. The merger was followed for 10 Gyr using the GADGET-2 $N$-body code (Springel,
Yoshida \& White 2001; Springel 2005) and we saved 201 simulation outputs, one every 0.05 Gyr.

\begin{figure}
\begin{center}
    \leavevmode
    \epsfxsize=7.6cm
    \epsfbox[0 20 260 135]{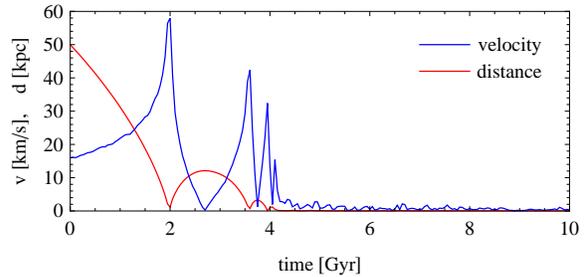}
\end{center}
\caption{The distance between the centres of the two merging dwarf galaxies (red line) and the relative
velocity between them (blue line) as a function of time.}
\label{distvel}
\end{figure}

The interaction between the two galaxies proceeded as illustrated in Figure~\ref{distvel} where the red line
shows the distance between the centres of the two merging dwarfs and the blue line their relative velocity. The
first fly-by occurred after about 2 Gyr of evolution and the two galaxies merged completely soon after 4 Gyr since
the start of the simulation.

\vspace{-0.5cm}

\section{Properties of the merger remnant}

The stellar component of the merger remnant is triaxial and retains some of the rotation of the initial disks.
Interestingly, this rotation is around the major axis of the stellar component, as required by the data for And II.
Such a result is due to the symmetry of the initial configuration: the components of the angular momentum vectors of
the two dwarfs along the $X$ axis of the simulation box (see Figure~\ref{merger}) were almost
equal and preserved during the
merger while the components of the angular momentum along the $Z$ axis of the simulation box had opposite signs and
cancelled each other.

\begin{figure*}
\begin{center}
    \leavevmode
    \epsfxsize=10.9cm
    \epsfbox[0 12 450 415]{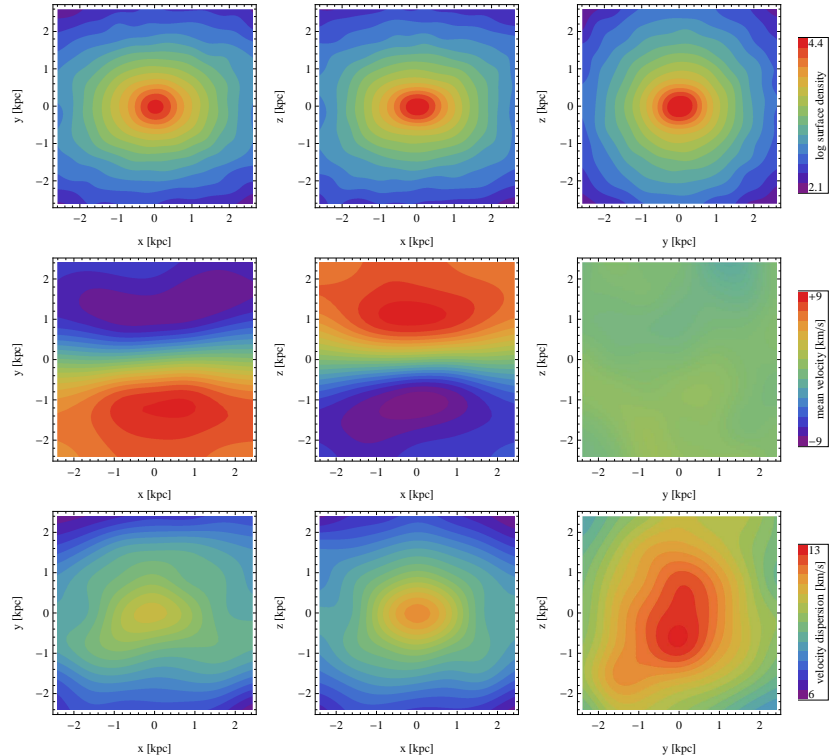}
\end{center}
\caption{Line-of-sight properties of the merger remnant. The upper, middle and lower row show the surface density
distribution of the stars, their mean velocity and velocity dispersion as would be measured by a distant observer.
The left, middle and right column correspond to the line of sight along the shortest ($z$), intermediate ($y$) and
longest ($x$) axis of the stellar component.}
\label{observed}
\end{figure*}

\begin{figure}
\begin{center}
    \leavevmode
    \epsfxsize=6.8cm
    \epsfbox[3 10 185 357]{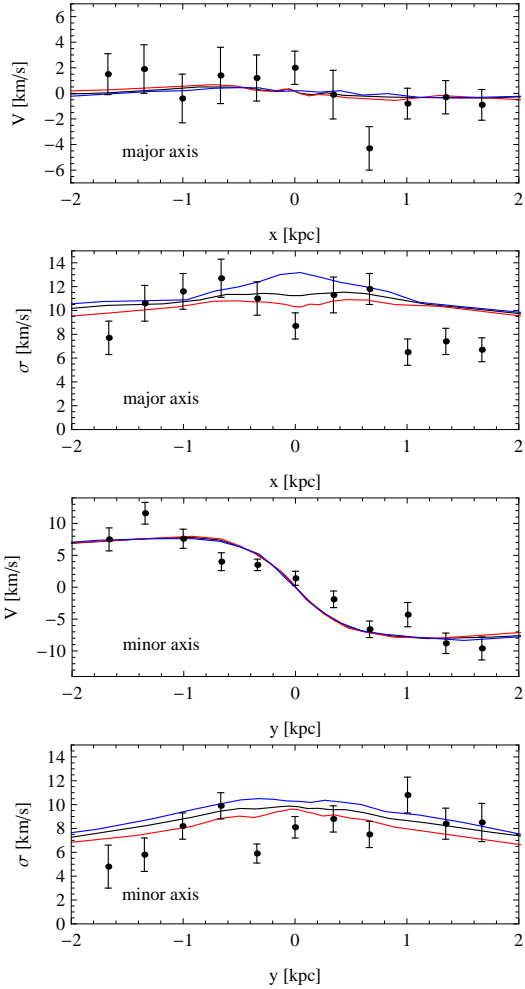}
\end{center}
\caption{Comparison between the kinematic profiles predicted by the model and real data. The line of sight chosen for
this comparison is the one along the shortest ($z$) axis of the stellar component
(left column plots of Figure~\ref{observed}).
The two upper (lower) panels show the mean velocity and velocity dispersion of the stars measured along the
observed major (minor)
axis of the galaxy image. Points with error bars are real data for all stars
in And II from Ho et al. (2012). Lines plot measurements from the simulation with black lines showing the measurements
for all stars while red (blue) ones for stars originating from Dwarf 1 (Dwarf 2).}
\label{veldispprof}
\end{figure}

The line-of-sight properties of the resulting dSph galaxy are shown in Figure~\ref{observed}.
We verified by a least-square fit that a very good match to the real data is obtained when the
line of sight is along the shortest ($z$) axis of the stellar component (left column
plots of Figure~\ref{observed}). In this configuration, the major and minor axis of the stellar image correspond to
the $x$ and $y$ (longest and intermediate) axes determined from the combined distribution of stars from both dwarfs
in 3D. The stars in the
simulated galaxy were binned along the major and minor axis of the image and the measured kinematic properties are
plotted as solid lines in Figure~\ref{veldispprof}. The black lines correspond to measurements for all stars from the
two dwarfs combined while the red and blue lines correspond to measurements for stars originating from Dwarf 1 and
Dwarf 2 respectively.

On top of these we plot as points with error bars the kinematic
measurements for And II from Ho et al. (2012) for all stars assuming the distance to And II of 652 kpc
(McConnachie et al. 2005).
The Figure demonstrates that the kinematics predicted by our model is a good match to the real data.
In particular, the model reproduces the lack of rotation along the major axis and a significant amount of rotation
along the minor axis. The decreasing dispersion profiles are also in reasonably good agreement with the data.
In addition, the kinematics of the two components are not very different from those measured for all stars.
We note that even at the end of the evolution, a few Gyr after the merger, there is a significant amount of
substructure present in the merger remnant in the form of streams and shells. These are not clearly visible in
Figures~\ref{observed} and \ref{veldispprof} because of the binning applied to the simulation data,
except for the irregularities in the velocity dispersion maps, similar to those found by
Amorisco et al. (2014) in the real data.

Our model also provides a reasonably good match to the data in terms of the photometric properties of the dwarf galaxy.
The ellipticity measured at the projected radius of $\sim1$ kpc for the line of sight along the shortest axis
(upper left panel of Figure~\ref{observed}) is of the order of 0.23 in good agreement with the determination in
Ho et al. (2012, their figure 2) and McConnachie \& Irwin (2006). The surface density profile of the stars can be
approximated by an exponential profile with a scale-length of 0.61 kpc, close to the scale of 0.67 kpc estimated
for And II by McConnachie \& Irwin (2006). Finally, adopting a mass-to-light ratio
of 4 M$_{\odot}$/L$_{\odot}$ for the stellar component in the visual band (within the range appropriate for old
stellar populations, see e.g. Bruzual \& Charlot 2003) our total stellar mass of $4 \times 10^7$ M$_{\odot}$
translates into the total luminosity of $10^7$ L$_{\odot}$, close to the real luminosity of And II,
$L_V = 9.6 \times 10^6$ L$_{\odot}$ (McConnachie \& Irwin 2006).

Da Costa et al. (2000) and McConnachie, Arimoto \& Irwin (2007) identified a few different stellar populations in And II
that can roughly be divided into
two: a metal-rich, intermediate-age and more centrally concentrated population and a metal-poor, old and more extended
one. A similar rough division was adopted by Ho et al. (2012) in their study of the kinematics of the different
populations (see their section 5.4, figures 9-10). Although we cannot attempt to reproduce all the intricacies of
stellar population properties with our simple collisionless model, we attempt a rough comparison with the data
also in this respect. We may identify our Dwarf 1 as the origin of the centrally concentrated population and Dwarf 2
as giving rise to the more extended one.

\begin{figure}
\begin{center}
    \leavevmode
    \epsfxsize=6cm
    \epsfbox[5 15 215 215]{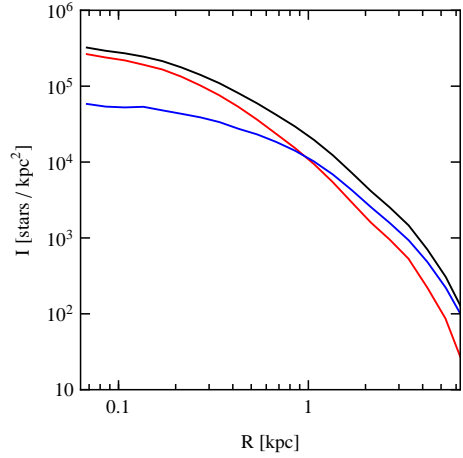}
\end{center}
\caption{Surface number density profiles measured in circular shells for the merger remnant seen along the
shortest principal axis ($z$). The red, blue and black lines correspond to the stars originating from
Dwarf 1, Dwarf 2 and the total stellar content, respectively.}
\label{numdenprof}
\end{figure}

Figure~\ref{numdenprof} confirms that the stellar density profiles of the
stars originating from these dwarfs indeed are different. In particular, the stars showing the more
concentrated profile (red line in Figure~\ref{numdenprof}) outnumber those of the more extended one by a factor of
$\sim3$ near the centre and the two components are equally numerous at about 1 kpc (which corresponds to about 5 arcmin),
exactly as required by the data (see figure 9 in Ho et al. 2012). We emphasize that despite these differences in
number density profiles, the two components show very similar kinematics, as demonstrated in Figure~\ref{veldispprof}.
Note that such a scenario does not imply the presence of a bimodal metallicity distribution in And II.
Instead, the combination
of the two populations (aided perhaps by additional merger-induced star formation) may well explain the origin of
the metallicity gradient in And II recently discussed by Ho et al. (2014).

\section{Why tidal stirring would not work}

As mentioned in the Introduction, the tidal stirring scenario for the formation of dSphs offers a simple way to
explain the origin of rotation in these systems as the remnant rotation of the initial disks.
In order to compare the types of rotation occurring as a result of a merger and of the process of tidal stirring
we measured the mean rotation velocity around the minor, intermediate and major axis of the stellar component of
one of our dwarf galaxies (Dwarf 1)
undergoing a merger and compared them to similar quantities measured for the same dwarf
undergoing tidal evolution around a Milky Way-like host
(for the latter case the set-up in terms of the dwarf galaxy, the
host and the orbit is the same as in {\L}okas et al. 2014 except that a smaller number of particles was used
here).

\begin{figure}
\begin{center}
    \leavevmode
    \epsfxsize=7.4cm
    \epsfbox[5 10 185 185]{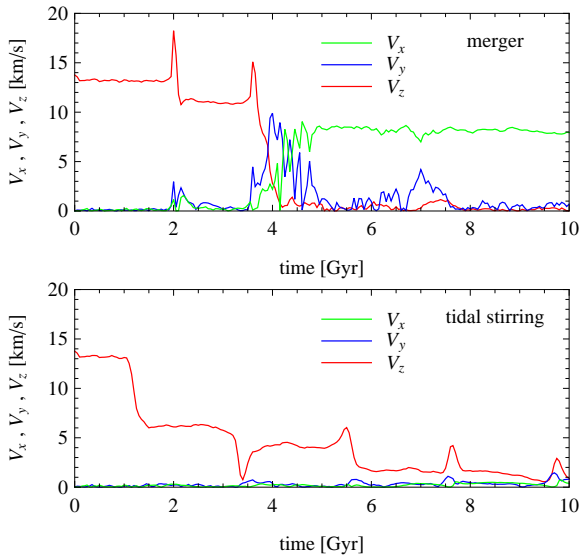}
\end{center}
\caption{The evolution of the mean rotation velocities around the major ($V_x$, green line), intermediate
($V_y$, blue line) and minor ($V_z$, red line) axis of the stellar component in the case of a dwarf galaxy
undergoing a merger (upper panel) and tidally stirred while orbiting a Milky Way-like host (lower panel).}
\label{rotations}
\end{figure}

The results of these measurements as a function of time are illustrated in Figure~\ref{rotations}.
The three mean rotation components were measured for stars within 1.5 kpc from the centre of the dwarf galaxy.
Initially the dwarfs have only rotation around the minor axis (of the order of 13 km s$^{-1}$) as expected for
disks. After the merger (upper panel of Figure~\ref{rotations}), the dwarf acquires rotation around the major axis
(of the order of 8 km s$^{-1}$), which is retained until the end of the evolution, while the remaining rotation components
become negligible. The situation is very different for a tidally stirred dwarf: its rotation around the minor axis
is diminished, especially at pericentres where sudden drops in its value occur, but no significant rotation around
the remaining axes is induced.

Note, that the example of tidal stirring discussed here is for one special configuration of the dwarf galaxy disk
with respect to the orbit
when it is exactly prograde. One could imagine that a different disk inclination could lead to the occurrence of
other rotation components. However, we have checked other inclinations of the disk and the result was always the same:
even if some rotation around the major or intermediate axis was induced, it was comparable or smaller than the remnant
rotation around the minor axis and always significantly smaller than the value of 1D velocity dispersion of the stars
(one such example is the simulation A discussed by {\L}okas et al. 2010).
Note that in the case of And II the rotation velocity in the outer parts of the dwarf is of the same order as the
velocity dispersion which makes it extremely difficult to explain by tidal evolution.

Another argument that makes the tidal evolution scenario improbable in the case of And II is related to its current
position with respect to M31. While the association between And II and M31 is rather beyond doubt in the sense that
the former is a satellite of the latter, And II is at present at a distance of 184  kpc from its host (McConnachie
2012). Even if this was
the maximum distance (apocentre of And II's orbit around M31) and the dwarf's orbit was very eccentric with a small
pericentric distance, it is unlikely that And II experienced more than two close encounters with M31 as the orbital
period in this case would be of the order of at least 4 Gyr (see the discussion of orbit O7 in {\L}okas et al. 2011).
Two pericentric passages could transform And II into a dSph galaxy only if the pericentric distance was very small
(of the order of 10 kpc).

\vspace{-0.5cm}

\section{Conclusion}

We presented a simple model for the formation of one of the dwarf spheroidal galaxies of the Local Group, And II,
that naturally explains the origin of the prolate rotation in this system. The dwarf galaxy with properties
akin to And II can be formed by a merger of two similar disky dwarf galaxies whose angular momenta are inclined
with respect to each other by 90 deg and collide on a purely radial orbit. The angular momentum of each disk
along the direction of the merger is preserved leading to the presence of prolate rotation in the final triaxial
merger remnant. The model accounts reasonably well for the observed photometric, and most importantly kinematic
properties of And II, reproducing in particular the amount of rotation versus random motion and their radial profiles.

We conclude that although tidal origin of the velocity distribution in And II cannot be excluded, it
is much more naturally explained within the scenario involving a past merger event. Thus, in principle, the presence of
prolate rotation in dSph galaxies of the Local Group and beyond may be used as an indicator of major
mergers in their history or even as a way to distinguish between the two scenarios of their formation.

An obvious extension of the present model would be to consider a merger where at least one dwarf enters the interaction
with a significant gas fraction. As in the well-studied case of mergers between big galaxies that lead to the formation
of giant ellipticals, one could then expect the gas to sink towards the centre and form new stars there (see e.g.
Cox et al. 2006). This would
naturally lead to the presence of multiple populations, with a younger one more concentrated towards the centre, but
would also explain the excess of stars in the inner part of the overall stellar profile of And II discussed
by McConnachie \& Irwin (2006, their figure 4). This picture is supported by a recent measurement of the star
formation history in And II (Weisz et al. 2014) showing an enhanced interval of star formation about 10-5 Gyr ago,
which can be identified with the time of the merger. We plan to address these remaining issues in our future work.

\vspace{-0.5cm}

\section*{Acknowledgements}

We thank L. Widrow for providing procedures to generate $N$-body models
of galaxies for initial conditions.
This research was supported in part by PL-Grid Infrastructure and
by the Polish National Science Centre under grant 2013/10/A/ST9/00023.
The work of IE was partially supported by the EU grant GLORIA (FP-7 Capacities; No. 283783).
IE and AdP are grateful for the hospitality of the Copernicus Center in
Warsaw during their visits. MS acknowledges the summer student program of the Copernicus Center.

\end{document}